
\input  phyzzx
\input epsf
\overfullrule=0pt
\hsize=6.5truein
\vsize=9.0truein
\voffset=-0.1truein
\hoffset=-0.1truein

%
%

\def\IC{{\ \hbox{{\rm I}\kern-.6em\hbox{\bf C}}}}
\def\IR{{\hbox{{\rm I}\kern-.2em\hbox{\rm R}}}}
\def\IZ{{\hbox{{\rm Z}\kern-.4em\hbox{\rm Z}}}}
\def\lcf{light cone frame}
\def\pa{\partial}
\def\sIR{{\hbox{{\sevenrm I}\kern-.2em\hbox{\sevenrm R}}}}

%
%
\hyphenation{Min-kow-ski}

\rightline{SU-ITP-95-23}
\rightline{November 95}
\rightline{hep-th/ 95}

\vfill

%
%
\title{Particle Growth and BPS Saturated States}

\vfill

%
%
\author{Leonard Susskind\foot{susskind@dormouse.stanford.edu}}

\vfill

\address{Department of Physics \break Stanford University, Stanford, CA
94305-4060}

\vfill

%
%
Consistency of the Bekenstein bound on entropy requires the physical dimensions
of particles
to grow with momentum as the particle is boosted to transplanckian energies. In
this paper the
problem of particle growth in heterotic string theory is mapped into a problem
involving the
properties of BPS saturated black holes as the charge is increased. Explicit
calculation based
on the black hole solutions of Sen are shown to lead to a growth pattern
consistent with the
holographic speculation described in earlier work.
%
\vfill\endpage

%
%

\REF\holo{L.~Susskind,{\it The World as a Hologram}, hep-th/9409089 }

\Ref\inf{L. ~Susskind, {\sl Infinite Momentum Frames and Particle Dynamics},
Lectures in Theoretical Physics. XID, Bolder Colo(1968)}

\REF\lorentz{L.~Susskind \journal Phys. Rev. & D49 (94) 6606 }

\REF\sen{A.~Sen, \journal Mod. Phys. Lett. & A10 (95) 2081, hep-th/9504147.}

\REF\Lennyone{L.~Susskind, {\it Some Speculations About Black Hole Entropy in
String Theory,}
Rutgers University preprint RU-93-44, August 1993, hep-th/9309145.}

\REF\STU{L.~Susskind, L.~Thorlacius, and J.~Uglum \journal Phys. Rev. & D48
(93) 3743.}

\REF\joe{J.~Polchinski, {\it Dirichlet-Branes And Ramond-Ramond Charges},
hep-th/9510017}

%
%

%
%
\chapter{Introduction}

According to the holographic principle of 't Hooft and the author, the physical
dimensions of
a transplanckian particle must depend on the relative motion of the particle
and observer in an
unconventional manner. In particular, the transverse cross sectional area must
grow at least
as fast as the longitudinal momentum. In free string theory, logarithmic growth
has been
known since the earliest days of string theory. Arguments were given in
[\holo] that
nonperturbative effects would cause the growth to accelerate to the more rapid
behavior. The purpose of this paper is to present additional evidence that this
is so.

The speculations of [\holo] can be summarized as follows:

1) Strings can be described by Hamiltonian quantum mechanics in the \lcf .

2) Strings are composed of
"string-bits". Each string-bit is a segment of string of length $\sqrt
{\alpha'}$. The information
content of a string never exceeds one bit per string-bit.

3) The transverse density of string bits in the light cone frame never exceeds
one string-bit per
Planck area. In what follows we will set G=1 so that the planck length is
unity.

4) A system of longitudinal momentum $P_-$ may be described as a collection of
string-bits of
total length $\alpha' P_-$ and therefore contains $\sqrt{\alpha'}P_-$
string-bits.

Evidently, the transverse area occupied by such a system must be at least

$$
A_{min} \approx {P_-}\sqrt{\alpha'}
\eqn\Amin
$$

The precise framework for this discussion is the \lcf. In this gauge,
relativistic mechanics
has a galilean  structure, [\inf] , which allows concepts of particle size,
wave functions of
composites  and other similar nonrelativistic concepts to be formulated.
Furthermore, the \lcf\
is is the only known hamiltonian formulation of string theory. Light cone
coordinates are
defined by the line element

$$
ds^2 = g_{+-}dx^+ dx^-  + g_{+i}dx^+ dx^i +  g_{ij} dx^i dx^j
\eqn\litecone
$$
We assume that spacetime is flat at infinity. In this situation we may further
fix the gauge so
that $g_{-+} =1$, $g_{+i} = 0$ and $g_{ij}=-\delta_{ij}$ as $X_- \to \infty$.
It is easy to see
that the trajectories $X^i = const$ , $X^+ = const$ are light like geodesics.

In these coordinates the transverse spread of a particle ${R_\perp}^2$ should
grow like

$$
{R_\perp}^2 \to P_-
\eqn\gro
$$

The symmetries of the light cone formulation include various kinematical or
nondynamical
symmetries such as transverse rotation and translation which are realized in a
trivial manner.
Among these are the transverse galilean  boosts under which each transverse
momentum $P_\perp$is
shifted by an amount proportional to the corresponding longitudinal momentum
$P_-$

$$
P_\perp \to P_\perp + v P_-
\eqn\boost
$$
Under these symmetries, transverse dimensions are unchanged.

In the \lcf\ the role of time is played by $X^+$. The initial data surfaces
$X^+ = const$ are
composed of light rays. The transverse coordinate of an event $p$ may be
obtained  as follows:

Choose a $2+1$ dimensional surface at spatial infinity defined by

$$
X^+ + X^- = L
\eqn\screen
$$
where $L$ is a large constant which eventually tends to infinity. In Ref.\holo\
this surface
was called "the screen". Assuming flat conditions at infinity, the screen may
be equipped
with cartesian coordinates with $g_{ij} = -\delta_{ij}$.  The transverse light
cone coordinates
of $p$ are found by passing a light ray through $p$ which intersects the screen
at right
angles. The transverse coordinates of this "image" point on the screen are also
the transverse
coordinates $X_\perp(p)$.  It will not be important in what follows to define
the longitudinal
coordinate of $p$. The statement that a particle grows with $P_-$ should be
taken to mean that
its image on the distant screen grows.

%
%
\chapter{Particle Size and BPS Black Holes}

We will consider heterotic string theory compactified on a 6 torus. The four
noncompact
dimensions are $X^{\mu} = (X^+,X^-, X^i)$ where $i= 1,2$. We also use the
notation $X^{0}=({X^+}
+{X^-})$, $ X^3=({X^+}-{X^-}) $, and $X^m=(X^1,X^2,X^3)$. One of the six
compact
coordinates will be singled out and called $Y$. The others play no role and
will be ignored. The
compactification radius for $Y$ is called $R_c$ and is assumed to be larger
than the fundamental
string length
$l_s = \sqrt {\alpha'}$. In particular $R_c$ is assumed large enough to easily
contain a
fundamental particle such as a graviton.

Let us consider a state with such a particle with vanishing transverse momentum
and
longitudinal momentum $P_-$. We wish to know if $R_\perp^2(P_-)$ increases with
$P_-$ and if so
how. To answer this, let us make a transverse galilean boost along the $Y$
direction until
$P_y =P_-$. This should have no effect on the size. The result of such a
transverse boost is to
bring the particle to rest in the $X_3$ direction. To see this note that $P_+ =
({P_\perp}^2 + {P_y}^2)/{P_-}$. Therefore in the present case $P_+=P_-$ and
therefore, $P_3 =0$.

The object we have constructed can be described in another way. It is a Kaluza
Klein charged
particle with charge $Q=P_y =P_-$ and mass equal to its charge. In heterotic
string theory it
is a BPS saturated particle with $Q_L=Q_R$ [\sen].

Let us first consider the transverse size of these objects in the free string
approximation. By
T-duality we can map the system to one which is compactified on a circle of
radius
${R_c'}={{\alpha'}\over {R_c}^2}$ with winding number $P_- R_c = N$. The total
length of the
wound string is
$N{R_c'} = P_- {\alpha'}$. It is well known that such a string will fluctuate
in the
orthogonal directions giving a growth proportional to $\log(P_- {\alpha'}) $.
Thus we recover
the usual logarithmic growth of weakly coupled strings. However this
calculation is inadequate
when the winding number becomes large. In this case, no matter how small the
string coupling is
the number of overlapping strings passing through the same place will be so
large that
interactions can not be ignored. Fortunately, in this limit, a semiclassical
description of
these objects is available.  For large $Q$ these objects are considered to be
extreme black holes whose large distance behavior  are described by classical
solutions of low
energy field theory. The classical solutions have been given by Sen [\sen] .

%
%
\chapter{Properties of Sen's Black holes}

In the limit of large mass and charge, these extreme BPS objects are described
by a metric with
the form

$$
ds^2 = g_{00} dt^2 -d{\rho}^2 - {\rho}^2 d{\omega}^2
=g_{00} dt^2 - dX^m dX^m
\eqn\senmet
$$
where $\rho$ is the proper distance from the horizon and

$$
g_{00} = {{\rho^4} \over {\rho^4 + M\rho^3}}
\eqn\goo
$$

The horizon at $\rho=0$ is singular. According to Sen [\sen] , the classical
solution should only
be believed for $\rho > \sqrt{\alpha'}$. At $\rho = \sqrt{\alpha'}$ a stringy
stretched horizon,
Ref \STU, \Lennyone\ is present. The stretched horizon contains whatever
distinctions exist
between different states. We will identify the region of the stretched horizon
as the object
whose growth we are interested in.

At first sight the situation seems disappointing. According to eq.\senmet\ the
area of the
stretched horizon is of order $\alpha'$ for all $M$ which if taken at face
value would say that
the dimensions are independent of $P_-$. However we are not yet in the light
cone frame. To
obtain the transverse spread in the \lcf\ we need to project the black hole to
the screen. This
requires solving for the light like geodesics.

%
%
\chapter{Projection to the Screen}

To find the light like geodesics in the background \senmet\ we use the usual
action principle.

$$
\delta \int \big[g_{\mu\nu}{dX^{\mu}\over d{\tau}} {dX^{\nu}\over
d{\tau}}\big]d\tau =0
\eqn\geodesic
$$

which gives

$$
\delta \int \big [g_{00}({dt\over d{\tau}})^2 - {dX^{m}\over
d{\tau}}{dX^{m}\over
d{\tau}}\big]d\tau =0
\eqn\geo
$$

The equation of motion for $t$ gives

$$
g_{00}{dt \over d\tau} = 1
\eqn\teq
$$
In eq.\teq we have arbitrarily normalized $\tau$ so that the right hand side is
unity.

The equations of motion for $X^m$ are

$$
2{{d^2X^m}\over{d\tau^2}} = -{\pa{g_{00}}\over{\pa X^m}}\big({dt\over
d\tau}\big)^2
$$.
Using eqn.\teq, this becomes

$$
{{d^2X^m}\over{d\tau^2}} ={1 \over 2} {\pa{g_{00}^{-1}}\over{\pa X^m}}
\eqn\xeq
$$
This is the equation of motion for a Newtonian particle in a potential

$$
V=-{1 \over{2 g_{00}}}
\eqn\poten
$$

The condition that the geodesic be light like reduces to the condition that the
total energy
vanish.

$$
{1 \over 2}\big[{dX \over d\tau}\big]^2 + V=0
\eqn\ll
$$

Using eq \goo we find that the potential $V$ is given by the familiar coulomb
potential plus an
additive constant.

$$
2V=-1 - {M\over \rho}
\eqn\coul
$$

Thus we see that light like geodesics correspond  to nonrelativistic particles
in a coulomb
potential which approach from infinity with unit velocity. The problem of the
size of the image
on the screen can now be solved. Assume that a light ray leaves the screen with
impact parameter
$b$ relative to the center of the coulomb potential. It will hit the stretched
horizon if $b$ is
less than some maximum value $R_{\perp}$. For a given $b$ the distance of
closest approach of the
trajectory can easily be obtained. It is given by the equation

$$
V(\rho) + {b^2 \over {\rho^2}} =0
\eqn\close
$$
If the distance of closest approach is less than $\sqrt{\alpha'}$ than the
trajectory hits the
stretched horizon and the original point is in image of the black hole. This
gives the size of
the image as

$$
R_{\perp}^2 = {\sqrt{\alpha'}} M =P_-{\alpha'}
\eqn\size
$$
for large $M$. Thus we see that the \lcf\ transverse size of a particle grows
with $P_-$ in the
required way.

%
%
\chapter{Two Particles}

In Ref.\holo\  a speculative picture was given for the behavior of a system of
well separated
particles moving with small relative motion when the combined system is boosted
to extreme
transplanckian momentum. The boost has no effect on the relative center of mass
energies of the
particles and should also have no effect on the invariant scattering
amplitudes. Consider, for
example, a pair of particles moving slowly past one another it the rest frame.
For simplicity we
assume that in this frame the particles move purely in the transverse
directions. Also assume they
pass one another at a  large macroscopic transverse distance $Z$ so that for
all
practical purposes they never interact.

Now boost the entire system along the longitudinal direction (orthogonal to
their motion). Each
particle grows so large that the image discs overlap as they pass. Nevertheless
they must
eventually emerge from the region of overlap with no scattering. This seems
quite remarkable.
We shall see that it is closely related to the very special properties of BPS
saturated states.

To analyze this system, again boost along the compact $Y$ direction until the
components $P_Y$
are equal to the longitudinal momenta. The system then consists of two BPS
black holes. Ignoring
the slow relative motion, we can write the metric for this system in the form
of eq.\senmet .
The only difference is that $g_{00}$ is now given by

$$
g_{00}^{-1} = 1 + {M \over |z-z_1|} + {M \over |z-z_2|}
\eqn\gooo
$$
For simplicity the two longitudinal momenta and therefore the BPS masses have
been chosen equal.
In eq.\gooo\ $z$ represents cartesian coordinates and $z_i$ are chosen so that
the images of the
black holes on the screen are centered at the  transverse locations of the two
particles in the
\lcf . Once again the motion of light rays is mapped into a nonrelativistic
coulomb problem,
this time with two force centers. The motion can be analyzed by elementary
methods.

If the separation between the transverse light cone locations of the centers of
the particles is large enough and $P_-$ is not too large, the
images will not overlap. We will also find that $|z_1 - z_2| > \sqrt
{\alpha'}$. As the mass
increases the individual images will grow  and for fixed distance between them
the value of $|z_1
- z_2|$ will decrease. Eventually, when $M$ becomes large enough  the images
will begin to overlap. At this point $|z_1 - z_2|$ will be about $\sqrt
{\alpha'}$.
In other words the stretched horizons begin to touch. Beyond this the images
presumably merge
into a single blob.

We are considering a system of particles which are slowly moving and well
separated in the rest frame. How, from the boosted viewpoint, can we understand
the fact that
the overlapping objects  do not interact? The answer lies in the very special
properties of BPS
states. Since the particles were boosted in the same direction along the $Y$
axis they are mapped
into same sign BPS particles. It is a well known fact that the forces between
such objects exactly
cancel if they are at rest no matter how close the are. The cancellation of the
long range
massless exchanges is easily seen from the low energy equations of motion.
However, much more is
involved. In particular cases such as D-branes [\joe] it can be seen that the
entire tower of
masses exchanges cancels exactly. In any case, for particles at rest, the
cancellation is thought
to be an exact consequence of supersymmetry.

On the other hand if the original particles are not at rest
the forces will not cancel. In particular, if the relative velocity between
them is large we
expect that they will undergo a violent collision, perhaps leading to  genuine
nonextremal black
hole formation. This is exactly what we expect of BPS particles if they collide
with appreciable
relative velocity.

\ack
I would like to thank Ed Witten for hospitality and for numerous interesting
discussions at the Institute for Advanced Studies where this work was carried
out.
I am also grateful to Renata Kallosh for many interesting discussions and
helpful explanations
of BPS black holes.
\refout
\end